\documentclass{article}

\usepackage{arxiv}

\usepackage[utf8]{inputenc} 
\usepackage[T1]{fontenc}    
\usepackage{hyperref}       
\usepackage{url}            
\usepackage{booktabs}       
\usepackage{amsfonts}       
\usepackage{nicefrac}       
\usepackage{microtype}      
\usepackage{lipsum}		
\usepackage{latexsym}  
\usepackage{amsmath, amssymb,amsthm}
\usepackage{algpseudocode}
\usepackage{tikz}
\usepackage{tabularx,colortbl}
\usepackage{comment}
\usetikzlibrary{arrows,automata}
\usepackage{bm}

\usepackage{fancyvrb}
\usepackage{listings}
\usepackage{color}
\usepackage{tikz}
\usetikzlibrary{arrows, positioning, automata, shapes, calc}

\definecolor{dkgreen}{rgb}{0,0.6,0}
\definecolor{gray}{rgb}{0.5,0.5,0.5}
\definecolor{mauve}{rgb}{0.58,0,0.82}

\usepackage{mathtools}
\usepackage{booktabs}
\usepackage{amsmath}
\usepackage{amssymb}

\usepackage{color}
\usepackage[symbol]{footmisc}
\usepackage[algoruled,vlined,linesnumbered]{algorithm2e}
\usepackage{units}
\usepackage{subcaption,booktabs}
\newtheorem{df}{Definition}

\newtheorem{col}{Corollary}
\newcommand{\bt}{\begin{theorem}\em}
\newcommand{\et}{\end{theorem}}
\newcommand{\bl}{\begin{lemma}\em}
\newcommand{\el}{\end{lemma}}
\newcommand{\bc}{\begin{col}\em}
\newcommand{\ec}{\end{col}}

\newcommand{\bea}{\begin{eqnarray}}
\newcommand{\eea}{\end{eqnarray}}
\newcommand{\bdf}{\begin{df}\em}
\newcommand{\edf}{\end{df}}
\newcommand{\ben}{\begin{enumerate}}
\newcommand{\een}{\end{enumerate}}

\title{An Evaluation Framework for Interactive Recommender Systems}

\author{
  Oznur Alkan\\
  IBM Research\\
 Dublin, Ireland\\
  \texttt{oalkan2@ie.ibm.com} \\
   \And
 Elizabeth M. Daly \\
   IBM Research\\
 Dublin, Ireland\\
  \texttt{elizabeth.daly@ie.ibm.com} \\
  \And
 Adi Botea \\
  IBM Research\\
 Dublin, Ireland\\
  \texttt{adibotea@ie.ibm.com} \\
}

\begin{document}
\maketitle

\begin{abstract}
Traditional recommender systems present a relatively static list of
recommendations to a user where the feedback is typically limited
to an accept/reject or a rating model. However, these simple modes
of feedback may only provide limited insights as to why a user likes
or dislikes an item and what aspects of the item the user has considered.
Interactive recommender systems present an opportunity
to engage the user in the process by allowing them to interact with
the recommendations, provide feedback and impact the results in
real-time. Evaluation of the impact of the user interaction typically
requires an extensive user study which is time consuming and gives
researchers limited opportunities to tune their solutions without
having to conduct multiple rounds of user feedback. Additionally,
user experience and design aspects can have a significant impact
on the user feedback which may result in not necessarily assessing
the quality of some of the underlying algorithmic decisions in the
overall solution. As a result, we present an evaluation framework
which aims to simulate the users interacting with the recommender.
We formulate metrics to evaluate the quality of the interactive recommenders
which are outputted by the framework once simulation
is completed. While simulation along is not sufficient to evaluate
a complete solution, the results can be useful to help researchers
tune their solution before moving to the user study stage.
\end{abstract}


\section{Introduction}
Most real-world recommender systems support very limited interaction where feedback and preferences are based on user item history and ratings or additional context information provided by the user which can become outdated~\cite{jannach2016user}. In order to overcome this challenge interactive recommender systems have gained attention by allowing users to have more control and making them more actively
involved in the process of their recommendations. One challenge
that has made it difficult for researchers to progress in this field
is that, there exists no standard evaluation mechanism for such
a framework. As a result, we have developed a sound evaluation
framework to simulate users as if they are interacting with the recommender.
The interaction we designed assumed to be performed
via conversation. The system can work with any recommender
algorithm that supports recommendations based on a user profile
and adapting the recommendations based on the learnt preferences.
In addition, we formulated metrics that combine measures from
two related areas of research, recommendation systems and dialogue
management, in order to evaluate the interactive recommender from
both recommender accuracy and dialogue quality perspectives.

The interactions considered during designing the framework are
as follows:
\begin{itemize}
\item Item feedback. A user can accept or reject an item.
\item Item exploration, explanations and correcting incorrect assumptions.
A user can examine item details or ask for explanation
for why an item is recommended. Once a user understands the
motivation behind a recommendation, then the user has the opportunity
to correct any incorrect assumptions. For example, the recommender may provide an explanation such as \textit{"I recommend this movie, because you prefer to watch Action and Fantasy"}. If the user does not actually like Fantasy, then the user may correct this via telling \textit{"But I do not prefer Fantasy"}.
\item Providing explicit preference. A user can provide preferences
(negative/positive/complex) towards features of items at any point
of conversation. The preference can be provided as a positive (e.g. I
like horror movies.) or negative (e.g. I do not like romantic comedies.)
preference towards a single feature, or a compound feature including
both a positive and a negative preference, (e.g. I want to watch
a movie of Keanu Reeves, but not a romantic one.)
\item Preference elicitation. The system can provide the user with a
choice of features to provide feedback on, to enrich the user preference
profile ( e.g. Do you prefer comedy movies?). The preference
elicitation implemented uses a feature selection method based on
Information Gain [2] to select the features that can split the recommendation
space better and allow the recommender to filter
out candidates. The output of the preference elicitation process is
therefore the feature whose value will be asked to the user.
\end{itemize}

Although the framework currently only supports the above interaction
mechanisms, the solution is generic enough to support
extensions to other interaction mechanisms as needed.
\section{Proposed Evaluation Framework}
\subsection{Overview}
In order to run the evaluation framework, any recommendation dataset that includes user-item ratings together with the item related data can be used. The data for each user is split into training and test, where the training items are used to generate the first set of recommendations and the test items, which is called as the \textit{look-ahead data}, is used to simulate the interaction. 
\begin{figure*}[t]
\centering
\includegraphics[width=0.63\textwidth, trim={6.5cm 3.2cm 3.3cm 2.5cm}]{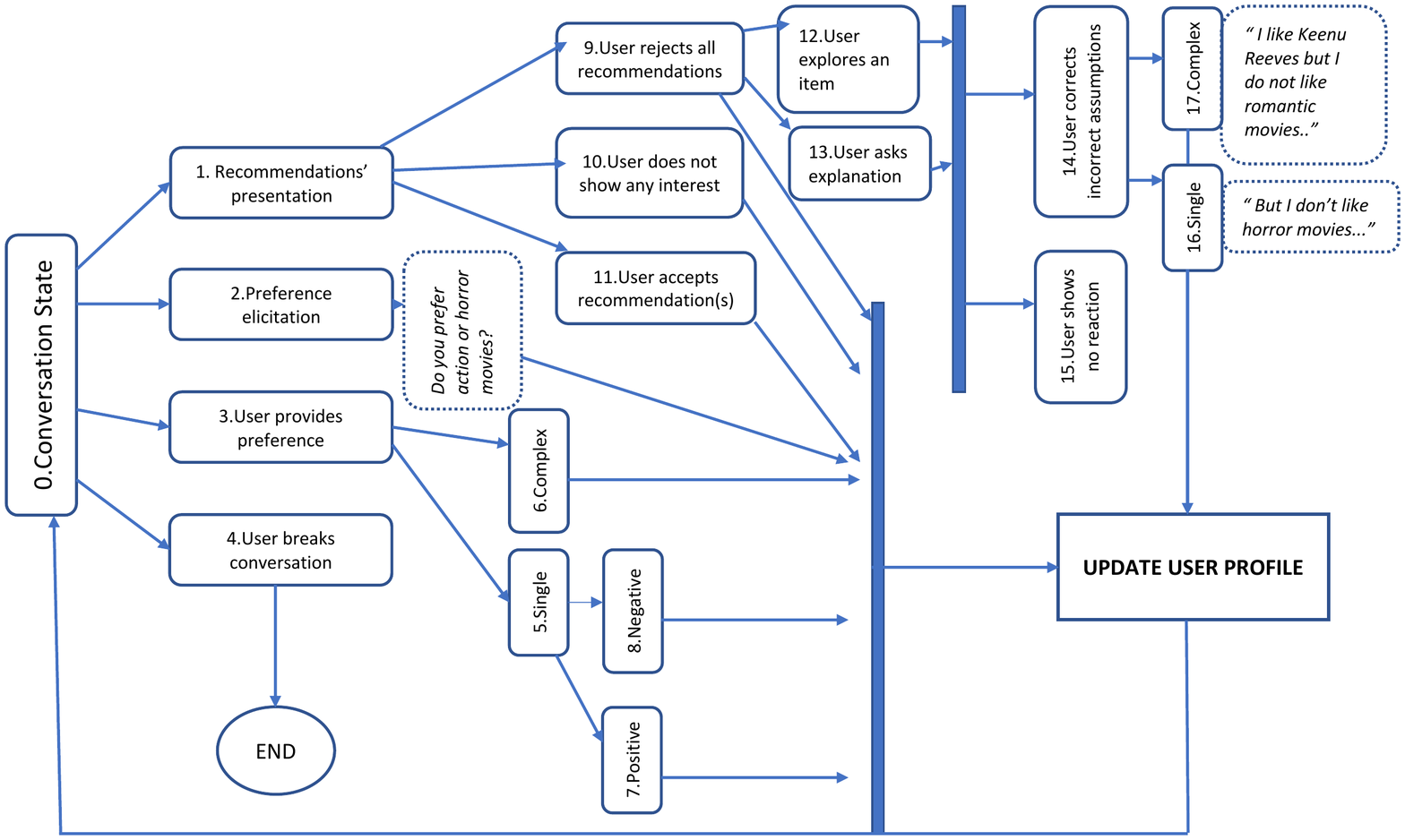}
\caption{User Simulation Outline}
\label{fig:experiment}
\end{figure*}

The simulation framework supports different \textit{conversation states} where the \textit{look-ahead data} is used to determine the simulated users responses for each conversational state.  
Each interaction is captured by a conversation state denoted by the nodes in Figure \ref{fig:experiment}. The \textit{look-ahead data} is used in order to determine the simulated users' choices at each state. The child nodes of each parent node reflect the possible next states that are chosen either by randomly or by leveraging the \textit{look-ahead data}. For example, when items are recommended, users may reject all of them or they may want to explore one of the recommended items to observe the properties of it and seek an explanation for why that item was recommended. Alternatively, the user may accept items, or they may provide explicit preferences based on the features of the recommended items. Without loss of generality and considering all the possible interactions users may experience with any conversational recommender, we created simulated users by following the interactions outlined in Figure \ref{fig:experiment}  and the underlying algorithm is summarized in Algorithms \ref{alg:evalMain}, \ref{alg:corrAssmp}, \ref{alg:providefeature}.

\subsection{Conversation States}
The goal is to simulate user interaction until the user breaks the conversation. Figure \ref{fig:experiment} shows the currently supported conversational states.
\\
\textbf{Interaction initialization state \textit{(state 0):}} At this point there are \textit{4} possible interactions; recommendations' presentation  \textit{(state 1)}, system asking a preference elicitation question \textit{(state 2)}, user providing explicit preference \textit{(state 3)}, and user breaking the conversation which ends the whole evaluation dialogue with a test user \textit{(state 4)}. Next state is chosen  in a weighted random manner. 
\\
\textbf{Recommendations' presentation \textit{(state 1)}:} The \textit{look-ahead data} is used to assess the user's reaction to the recommendations. If the user has negatively rated the presented items, then \textit{state 9} is selected. If the future items in the \textit{look-ahead data} have no rating for the presented items, then we infer that the user has no interest for the suggestions \textit{(state 10)}, whereas if the user has positively rated at least one recommended item,  \textit{state 11} is selected. 
\\
\textbf{Preference elicitation \textit{(state 2):}} The system generates a preference elicitation option for the user, and the option selected is determined based on the \textit{look-ahead data}, where the features associated with future items are used to infer the preferences of the user. 
Recommender can ask questions to elicit user preferences from dialogue which is illustrated by state 2. The feature being asked for eliciting preference is selected by choosing the feature with the maximum information gain among the recommendation items.
\\
\textbf{User providing preference \textit{(state 3)}:} In a similar manner to above, the \textit{look-ahead data} is used to examine the features of the future items, where we can generate a preference the user chooses to give to the system.  \textit{e.g. I dont like romantic movies.}.

For the state of providing complex preferences (state 6), if no negative features are found in the \textit{lookahead data}, to simulate user providing complex features, 
the simulation proceeds as follows: It checks whether there is any positively weighted feature in the current learnt profile that does not exist in the positively rated items of the \textit{look ahead data}. This corresponds to the scenario where users changed their mind for previously liked features. 

\subsection{Algorithmic Details}
The underlying algorithms used to evaluation framework are summarized in Algorithms \ref{alg:evalMain}, \ref{alg:corrAssmp}, \ref{alg:providefeature}.

Algorithm \ref{alg:evalMain} details the main algorithm behind the running the simulation framework. The goal of simulated user is to chat with the system until a point when the user wants to break the conversation. The boxes in Figure \ref{fig:experiment} corresponds to different conversation states that the system can be at during any point of the dialogue. The boxes are further numbered for the sake of easy referral within the paper. The box labelled as Conversation State, which is state 0, stands for the state where the evaluation framework selects between four possible interactions, namely, \textit{generating recommendations, system asking a preference elicitation question, user providing explicit preference,} and \textit{user breaking the conversation} which ends the whole evaluation dialogue with a test user, in a weighted random manner. This is illustrated with ProbRand() function at line 33 of Algorithm \ref{alg:evalMain}. For the evaluation, from each dataset \textit{80\%} of the users are randomly selected to be used in training the recommender, and remaining \textit{20\%} are used to evaluate the solution. For each test user in \textit{20\%}, first  \textit{80\%} of the movies are used to generate recommendations and remaining  \textit{20\%} is used to simulate the interaction, where this  \textit{20\%} stands for a look-ahead data for the evaluation framework, such that, we pretend as if users future choices are in these \textit{20\%}.

\begin{algorithm}
\small
\DontPrintSemicolon
\SetAlgoLined
\SetKwInOut{Input}{Input}\SetKwInOut{Output}{Output}
\Input{$testuser:<testset,$\;
            $trainset\{neg,pos,avg\_rtg\}$}
\BlankLine
$ignored  \gets  \{\}$, 
$s \leftarrow generate\_recs$ \;
\BlankLine
\While{\textit{true}}{    
        \uIf{$s = generate\_recs$} {
            $recList \gets$ \textbf{ \textsc{callRec(}}$testuser, ignored$\textbf{)} \;
            \uIf{$recList \cap testset$}{
                $acceptL \gets recList \cap pos$ \;
                \uIf{$acceptL \neq \O$}{
				   \textbf{ \textsc{updateProfile(}}$acceptL$\textbf{)}
                }
                \uElse {
				   \textbf{ \textsc{updateProfile(}}$recList$\textbf{)}\;
				   $s \gets$   \textbf{\textsc{rand(}}$\{explore,explain,no\_react\}$\textbf{)}
					\uIf{$state \neq no\_react$}{
                       $s \gets$  \textbf{ \textsc{rand(}}$\{no\_react,corr\_assmp\}$\textbf{)}} \;
                       \uIf{$state = corr\_assmp$}{
                       $fp,fn \gets$  \textbf{\textsc{corrAssmp(}}$testset,recList$\textbf{)}
                         \textbf{ \textsc{updateProfile(}}$fp,fn$\textbf{)}}
			   }	   
            } 
            \uElse{
                 add $recList$ to $ignored$
            }
        }
        \uElseIf{$s = provide\_pref$}{
            $s \gets$  \textbf{\textsc{rand(}}$\{pos,neg,complex\} $\textbf{)} \;
            $fp,fn \gets$ \textbf{\textsc{ProvideFeature(}}$s,testset$ \textbf{)} \;
            \uIf{s = pos} { \textbf{ \textsc{updateProfile(}}$fp$\textbf{)}} 
            \uIf{s = neg} { \textbf{ \textsc{updateProfile(}}$fn$\textbf{)}} 
            \uElse{ \textbf{ \textsc{updateProfile(}}$fp,fn$\textbf{)}}
        }
        \uElseIf{$s = pref\_elicitation$}{
            $f \gets$ \textbf{ \textsc{PrefElic(}}$test\_user, ignored$\textbf{)}\;
            \textbf{ \textsc{updateProfile(}}$f$\textbf{)}
        }
        \uElseIf{$s = break\_conversation$}{
            break
        }   
        }  
        $s \gets$  \textbf{ \textsc{probRand(}}$\{break\_conversation,generate\_recs,$ \;
 $provide\_pref,pref\_elicitation\}$ \textbf{)}  \;    
\caption{\textsc{Evaluation}MAIN}
\label{alg:evalMain}
\end{algorithm}

\begin{algorithm}
\small
\DontPrintSemicolon
\SetAlgoLined
\SetKwInOut{Input}{Input}\SetKwInOut{Output}{Output}
\Input{$testset,recList$}
\Output{$fp,fn$}
\BlankLine
$profile,avgFeatWeight \gets$ buildUserProfile$(test\_movies)$\;
$neg\_feat \gets \forall f \in profile|weight(f)<avgFeatWeight$ \;
$pos\_feat \gets \forall f \in profile|weight(f) \geq avgFeatWeight$ \;
sort $neg\_feat$ ascending wrt weight\;
sort $pos\_feat$ descending wrt weight\;
\ForEach{nf in $neg\_feat$}{
   \ForEach{m in $recList$}{
   \uIf{m has nf}
   {
   	 $fn \gets nf$\;
   	 \ForEach{pf in $pos\_feat$}{
   	 	 \uIf{m has pf} {
   	  		$fp \gets pf$\;
   	  		break\;
   	  	}
   	  }
   	  break
   }
   }
}
\caption{\textsc{corrAssmp}}
\label{alg:corrAssmp}
\end{algorithm}

\begin{algorithm}
\small
\DontPrintSemicolon
\SetAlgoLined
\SetKwInOut{Input}{Input}\SetKwInOut{Output}{Output}
\Input{$s,testset$}
\Output{$fp,fn$}
\BlankLine
$profile,avgFeatWeight \gets$ buildUserProfile$(test\_movies)$\;
$neg\_feat \gets \forall f \in profile|weight(f)<avgFeatWeight$ \;
$pos\_feat \gets \forall f \in profile|weight(f) \geq avgFeatWeight$ \;
$fn \gets min(neg\_feat)$, $fp \gets max(pos\_feat)$
\caption{\textsc{providefeature}}
\label{alg:providefeature}
\end{algorithm}

For each test user, the conversation starts with the system generating the recommendations (line 1 of Algorithm \ref{alg:evalMain}). For each next turn of the conversation, one of the four states are chosen on a weighted random manner, such that, each state is  assigned a probability value, where a probability value of 0 ensures that, corresponding state never gets selected and probability value of 1 ensures that, it is always being selected. Sum of the probabilities of four states adds upto 1. If the system generates recommendations (line 3 of Algorithm \ref{alg:evalMain}), the possible interactions of the user are accepting at least one of them (lines 7-8 of Algorithm \ref{alg:evalMain}), rejecting all of them (line 10-15 of Algorithm \ref{alg:evalMain}), or showing no reaction which simulates the fact that user does not show any interest at all (line 17 of Algorithm \ref{alg:evalMain}) to any of the suggestions. In order to simulate this, the evaluation framework checks the look-ahead data, which is the remaining test set of the test user, and observes whether there exist any intersection between the test set and the generated suggestions. Each accept or reject results in shrinking the test set with the corresponding items and adding them to the training set for that test user, which is performed by UpdateProfile() calls in Algorithm \ref{alg:evalMain}. 

In the case of reject, which is state 9 in Figure \ref{fig:experiment} user may either explore or seek explanation for an item, or may not provide any further feedback. If explores or observes the explanation, user may inspect an incorrect assumption that the recommender has made, so user may provide feedback which can be a single feature or a complex one. This reflects a use case scenario where the recommender may provide an explanation with the features it used to recommend that item, however, the user may inspect that, features being used positively by the recommender are not the ones user likes. For example, recommender may explain recommending movie \textit{November Rain} by saying \textit{"I recommend you November Rain because you like Keenu Reeves, Charlize Theoron, movies from US and Romantic movies"}. User may correct the recommender by responding with a complex answer involving a positive and negative feature, such as, \textit{"I like Keenu Reeves but I dont like Romantic movies"}. To simulate this, the evaluation framework follows the methodology outlined in Algorithm \ref{alg:corrAssmp}. It first extracts all the negative and positive features based on movies in test set (lines 2-5 in Algorithm \ref{alg:corrAssmp}), and loops over recommendations to find out whether there exists a movie that has one of those negative features (lines 6-17 in Algorithm \ref{alg:corrAssmp}). If so, it checks whether the movie contains a feature from users positive features from test set as well (lines 10-14 in Algorithm \ref{alg:corrAssmp}). If both negative and positive feature exists, it assumes the user responds with both positive and negative features, which leads to state 17. Otherwise it assumes user responds with just correcting the incorrectly assumed positive feature, which leads dialogue to enter state 16. 

User can provide explicit preference, as represented by state 3 in Figure \ref{fig:experiment}. The provided preference can be a single positive feature (state 7), a single negative feature (state 8) or a complex feature which involves both a positive and a negative feature (state 6). In order to simulate this, the evaluation framework uses the look-ahead data, and selects the most positive and negative feature, which is outlined in Algorithm \ref{alg:providefeature}. For each user, while the user profile is being built by the recommender, maximum, minimum and average feature weights are being kept as part of user profile. All features having weight less than the average feature weight are assumed to be negative (line 2 of Algorithm \ref{alg:providefeature}). Similar hold for the positive case. Most negative and positive features are being returned by the ProvideFeature() call of line 20 of Algorithm \ref{alg:evalMain}. These features are set to either to minimum or maximum feature weights kept in user profile for the negative and positive states accordingly by the UpdateProfile() call at lines 22, 24, and 26 of Algorithm \ref{alg:evalMain}.

Recommender can ask questions to elicit user preferences from dialogue. This is illustrated by state 2 of Figure \ref{fig:experiment}. The feature being asked for eliciting preference is performed by selecting the most informative feature among the recommendation items as explained in section 3.1 of the paper.

The Rand() calls at lines 11,12, and 19  of Algorithm \ref{alg:evalMain} randomly selects between the states, however, ProbRandom() at line 33 selects based on weights assigned to states 1-4 within the evaluation framework. 

\section{Metrics}
Typical recommender system metrics focus on the accuracy of a presented list, however, considering interactions with the user, we want to be able to capture what the overhead is on the user. For example, simply rating movies might take quite a number of recommendations to hit a successfull recommendation, whereas asking the user if they prefer \textit{drama} or \textit{horror} might lead to success with a smaller number of interactions. As a result, we designed a number of metrics to evaluate not only the \textit{accuracy} of the recommendations but also the amount of additional interactions between the user and the recommender needed, which reflects the quality of the interactions. 
\begin{itemize}
\item \textbf{DialogTurn (DT): } \#dialogue turns, where each utterance of the user and the recommender is counted as one turn. In other words, each individual turn per user and recommender is counted separately and added to DT. ex: user presenting a preference, recommender generating suggestions, etc. 
\item \textbf{DialogSuccessRate (DSR):} avg \#successfull recommendations presented to the user during per dialogue turn. 
\item \textbf{RecommendationTurn (RT): }  \#times recommendations are presented to user. DT increases even if user presents a preference and   
\item \textbf{RecommendationSuccessRate (RSR):} avg \#successful recommendations generated per one recommendation turn.
\item \textbf{AP@k: } average precision of top \textit{k} recommendations \cite{LIa2011ASI}, used in calculation of  \textit{AP@kDT}. \textit{AP@k} is calculated by looking at the ground truth, which reflects whether the user likes each item in reality as opposed to generated suggestions at specified ranks. Based on this, \textit{AP@k }is calculated by taking the average of precisions computed at each liked position in the top \i\textit{k} items of the user's ranked list. 
\item \textbf{AP@kDT:} avg normalized AP@k by the \#dialogue turns.
\item \textbf{AP@kRT:} avg normalized AP@k by recommendation turn.
\end{itemize}

We use \textit{RT} as a different metric from \textit{DT}, as we believe RT reflects the success of interactive recommenders better than DT, which is used to evaluate dialogue systems in general.
We integrated these metrics within the framework, and at the end of the evaluation with each test user, the average of the calculated metrics  over test users is presented. 
The evaluation framework outputs the values calculated for these metrics once the evaluation for each test user is completed.

\section{Examples of Conversations}

Sample execution logs of the evaluation framework are presented in the following Figure \ref{fig:example}  for two sample test users, where for the first one, user breaks the conversation sooner than the second one.

\begin{figure*}
	\centering
	\begin{subfigure}{1\textwidth} 
		\includegraphics[width=\textwidth]{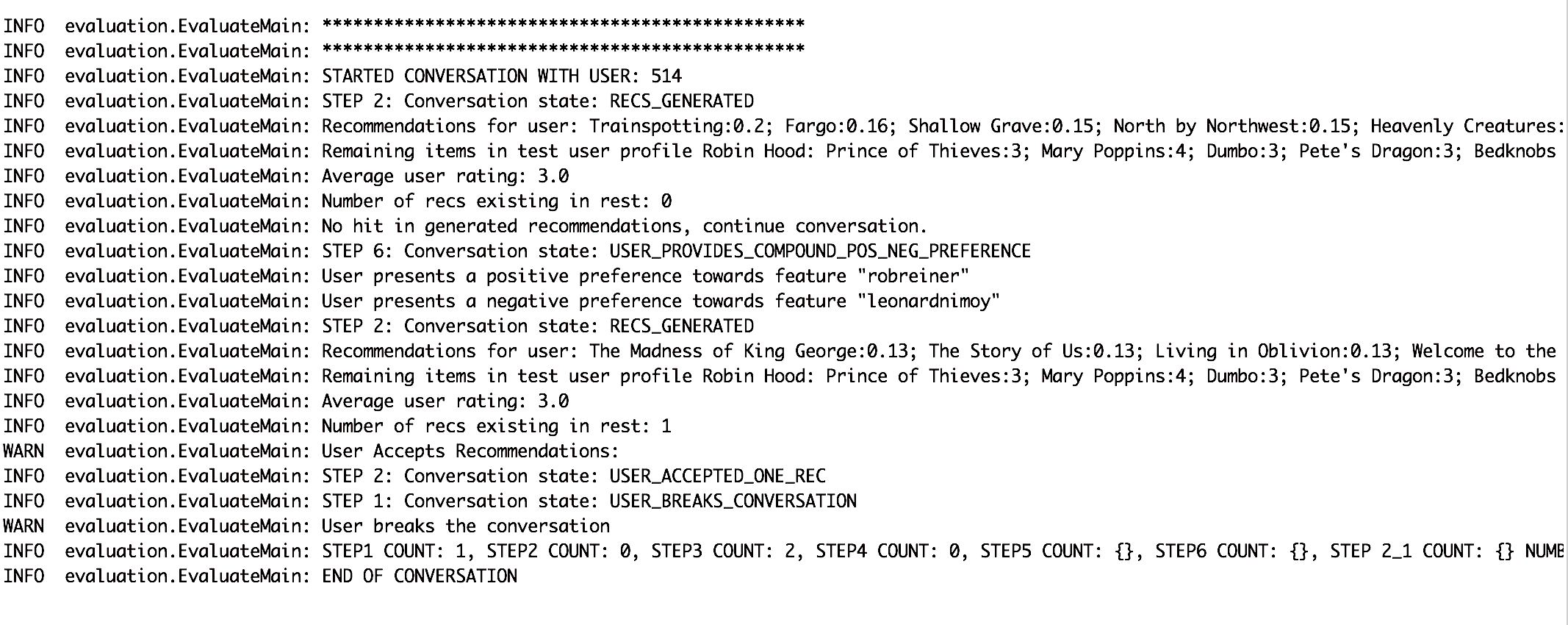}
		\caption{} 
	\end{subfigure}
	\vspace{1em} 
	\begin{subfigure}{1\textwidth} 
		\includegraphics[width=\textwidth]{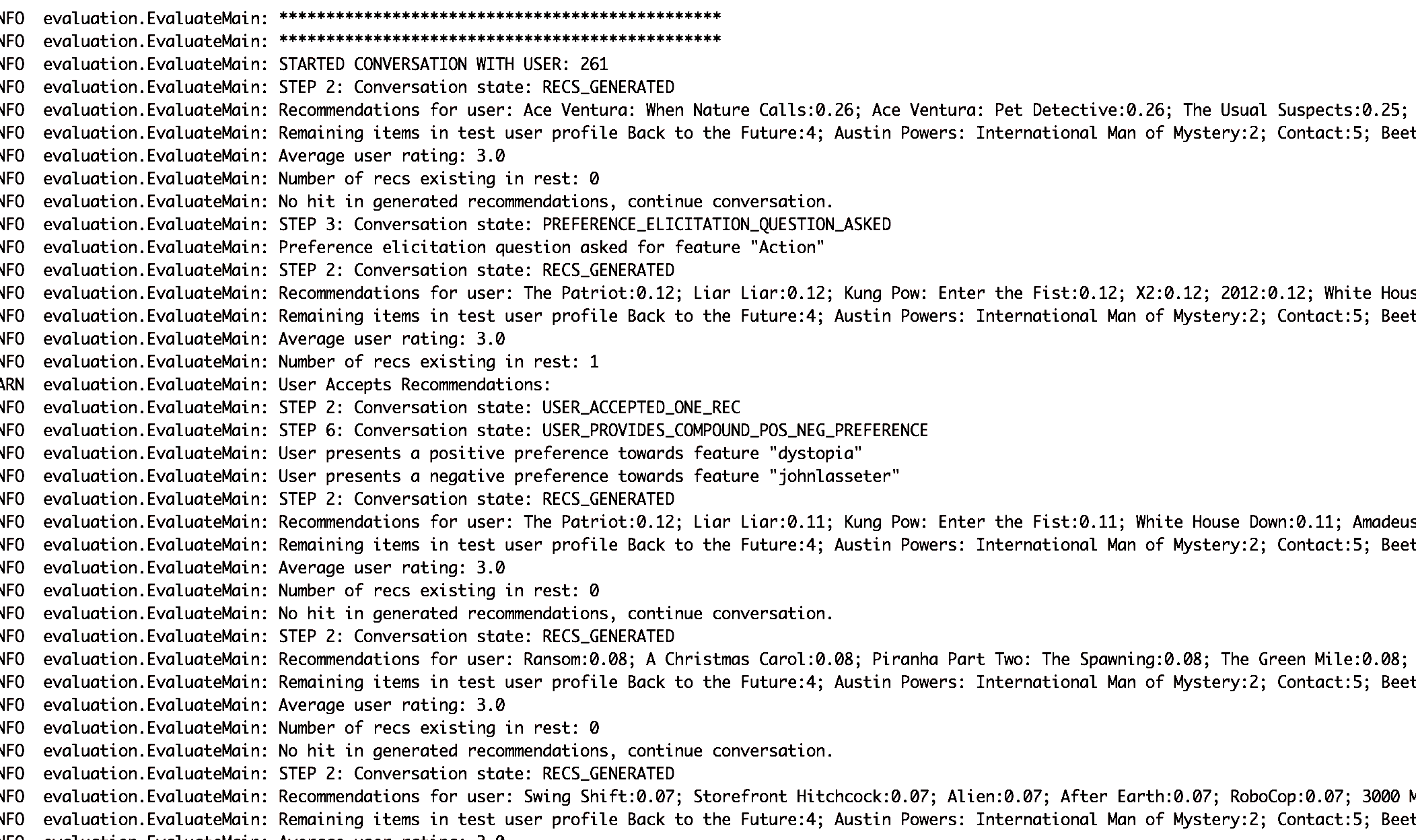}
		\caption{} 
	\end{subfigure}
		\caption{Conversation Examples from User Simulation Experiments} %
	\label{fig:example}
\end{figure*}

\bibliographystyle{unsrt}  
\bibliography{references}  

\begin{thebibliography}{1}

\bibitem{jannach2016user}
Dietmar Jannach, Sidra Naveed, and Michael Jugovac.
\newblock User control in recommender systems: Overview and interaction
  challenges.
\newblock In {\em International Conference on Electronic Commerce and Web
  Technologies}, pages 21--33. Springer, 2016.

\bibitem{LIa2011ASI}
Hang Li.
\newblock A short introduction to learning to rank.
\newblock 94-D:1854--1862, 10 2011.

\end{thebibliography}

\end{document}